\documentclass[english,keywords,showpacs,amsmath,amssymb,twocolumn]{revtex4}
\usepackage[T1]{fontenc}
\usepackage[latin1]{inputenc}
\usepackage{babel}
\usepackage{array}
\usepackage{graphics}
\usepackage{subfigure}
\usepackage{amsmath}
\makeatother
\begin{document}
\title{Reply to the ``comment on `Contextuality within quantum mechanics  manifested in subensemble mean values[Phys. Lett. A 373(2009)3430]' ''}
\author{Alok Kumar Pan\footnote{apan@bosemain.boseinst.ac.in}$^1$,
and Dipankar Home\footnote{dhome@bosemain.boseinst.ac.in}$^1$}

\affiliation{$^1$ CAPSS, Department of Physics, Bose Institute, Sector-V, Salt Lake, Calcutta
700091, India}

\begin{abstract}
In this reply, we point out that the claim by De Zela[2] is unjustified because the setup he discusses is \emph{not} equivalent to the setup analysed in our paper[1]. Hence his subsequent argument claiming the reproducibility of our demonstrated quantum effect of path-spin contextuality by the Kochen-Specker realist model is \emph{not} relevant.
\end{abstract} 
\maketitle
\section{Introduction}
In our earlier paper\cite{pan}, using a suitable variant of Mach-Zehender(MZ) interferometric setup,  we demonstrated a curious form of context-dependence \textit{within} the formalism of quantum mechanics(QM) between the path and the spin degrees of freedom  of a single neutron.  This effect of `path-spin contextuality' gets manifested at the level of subensemble mean values of a spin variable that are contingent upon which \emph{choice} has been  made for measuring an appropriately defined comeasurable(commuting)`path' observable. Our demonstration was with respect to the aforementioned setup involving three distinct stages; in the first stage, it was prepared a path-spin entangled state of single neutron, and subsequently used for measuring the path observable in the second stage, followed by the spin measurement in the final stage. Importantly, as expected, this effect gets obliterated at the level of quantum expectation value of the spin variable in question that is defined with respect has  to the entire ensemble of neutrons on which the measurements are performed. 

In a comment on our paper, De Zela claimed that our setup ``is nothing but a variant of the standard Stern-Gerlach(SG) setup'', and hence the type of path-spin context-independence  demonstrated in our paper can be successfully reproduced by the Kochen-Specker(KS) realist model for single qubit. In this paper, we show that his argument is based on a naively founded idea, because the variant of SG setup he has considered is by no means equivalent to setup that we had used in our demonstration. Hence, the question of reproducibility of the results of a variant of the SG setup by the KS model has no relevance to our argument. In fact, as we shall point out, in his setup no context(in terms of measuring the path observable) can be defined while considering the measurement of a spin variable which was the key ingredient in our scheme. 

Now, before proceeding further, in order to make our reply to be self-contained, we briefly recapitulate the essence of our earlier argument\cite{pan}. For clarity, we analyse our earlier setup by clearly pinpointing three different stages; the state preparation, the path measurement and finally the spin measurement. In contrast, the setup that has been used by De Zela which he has claimed to be equivalent to our setup  has two distinct stages; the state preparation and the spin measurement. A crucial point to note here is that the state prepared after the first stage of De Zela's setup is the state that is obtained in our setup after the two stages of state preparation and path measurement.
 
\section{Our earlier proposed experiment}

We considered an ensemble of neutrons, all corresponding to an initial spin polarized state along the $+\widehat {z}-axis$(denoted by $\left|\uparrow\right\rangle_{z}$) be incident on a  50:50 beam-splitter(BS1)(Fig.1). Next, the neutrons which move along one of the channels, say, the one corresponding to $|\psi_{1}\rangle$ pass through a spin-flipper(SF) that flips the state $\left|\uparrow\right\rangle_{z}$ to $\left|\downarrow\right\rangle_{z}$. Subsequently, the neutrons passing through the channels $\left|\psi_{1}\right\rangle$ and $\left|\psi_{2}\right\rangle$ are reflected by the mirrors M2 and M1 respectively - these reflections do not lead to any net relative phase shift between $\left|\psi_{1}\right\rangle$ and $\left|\psi_{2}\right\rangle$.

Thus, the state $|\Psi\rangle_{PH}$(it was $|\Psi\rangle$ in our earlier paper) of the neutrons incident on the second beam splitter(BS2) is represented by
\begin{eqnarray}
\left|\Psi\right\rangle_{PH}=\frac{1}{\sqrt{2}}\left(\left|\psi_{1}\right\rangle\left|\downarrow\right\rangle_{z}
+i \left|\psi_{2}\right\rangle\left|\uparrow\right\rangle_{z}\right)
\end{eqnarray}
The state given by Eq.(1) was our \emph{prepared state}(an \emph{entangled state} between the path the spin degrees of freedom of neutron) on which we had considered the path and the spin measurements relevant to our demonstration of contextuality. This was clearly mentioned in the Section 2 of our earlier paper\cite{pan}. But, in view of the comment by De Zela, we may again stress that the process of passing neutrons through the arrangement of BS1+SF serves the purpose of appropriately \emph{preparing} the state,  and the prepared state is subject to BS2 and two spatially separated Stern-Gerlach devices (SG1 and SG2)  for the path and the spin measurements respectively.   
\begin{figure}[h]
{\rotatebox{0}{\resizebox{9.0cm}{7.0cm}{\includegraphics{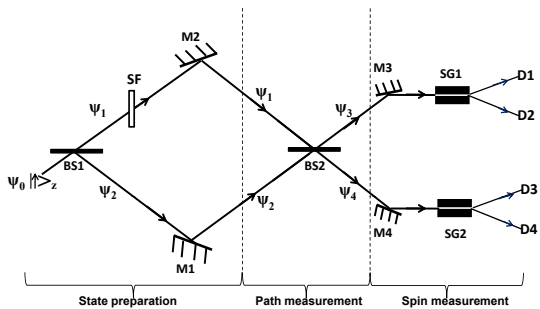}}}}
\vskip -1.2cm
\caption{\footnotesize The experimental setup used in our earlier paper\cite{pan}  is again depicted here that consists of a variant of Mach-Zehnder interferometer and two spatially separated Stern-Gerlach devices SG1 and SG2. This setup has three  distinct stages are the following: (i) The state preparation - the first beam splitter BS1 and and spin-flipper SF are used for preparing a path-spin entangled state (ii) The path measurement - the second beam-splitter BS2 plays this crucial role (iii) The spin measurement - which is performed by SG1 and SG2 corresponding to the channels $\left|\psi_{3}\right\rangle$ and $\left|\psi_{4}\right\rangle$ respectively. For inferring path-spin contextuality within quantum mechanics, \emph{subensemble mean values} of the spin variable $\widehat\sigma_{\theta}$ are considered pertaining to SG1 and SG2 separately.}
\end{figure}
\subsection{The path measurement}
For analysing the  relevant path measurement on the state $|\Psi\rangle_{PH}$( it was $|\Psi\rangle$in our earlier paper) which is prepared by the setup \emph{preceding}  BS2, we write the state after emerging from BS2 as given by
\begin{eqnarray}
\left|\Phi\right\rangle_{PH}= \frac{1}{\sqrt{2}}\left[i\left|\psi_{3}\right\rangle(\gamma \left|\downarrow\right\rangle_{z} + \delta   \left|\uparrow\right\rangle_{z})
+  \left|\psi_{4}\right\rangle\left(\delta\left|\downarrow\right\rangle_{z}-\gamma \left|\uparrow\right\rangle_{z}\right)\right]
\end{eqnarray}
where the output states $\left|\psi_{3}\right\rangle=-i\gamma \left|\psi_{1}\right\rangle+ \delta  \left|\psi_{2}\right\rangle$ and $\left|\psi_{4}\right\rangle=\delta\left|\psi_{1}\right\rangle - i\gamma\left|\psi_{2}\right\rangle$ with  $\gamma^{2}+\delta^{2}=1$. Then, for a given linear combination of  $\left|\psi_{1}\right\rangle$ and $\left|\psi_{2}\right\rangle$, using the different values of  $\gamma$($\delta$), one can generate at the output of BS2, various linear combinations of $\left|\psi_{1}\right\rangle$ and $\left|\psi_{2}\right\rangle$ that correspond to different probability amplitudes of finding particles in the channels corresponding to $\left|\psi_{3}\right\rangle$ and $\left|\psi_{4}\right\rangle$. Then, the different values of the parameter $\gamma(\delta)$ of BS2 can be regarded as a different \emph{choices} the `path' observables denoted by  $\widehat{A}_{\gamma}=P(\psi_{3})-P(\psi_{4})$ with the eigenvalues $+1(-1)$ correspond to $\left|\psi_{3}\right\rangle(\left|\psi_{4}\right\rangle)$ respectively.  Thus, BS2 plays a crucial role as a part of this measuring arrangement.

By using suitable representation of $\left|\psi_{1}\right\rangle$ and $\left|\psi_{2}\right\rangle$, the path observable can be written as 
\begin{equation}
\widehat{A}_{\gamma}= \left(\begin{array}{cl} 
    \gamma^{2}-\delta^{2}& \ -i 2\gamma\delta \\  i 2\gamma\delta  & \ \delta^{2}- \gamma^{2} \end{array}\right) 
\end{equation} 
Note that, different values of the parameter $\gamma(\delta)$ involving BS2 represents different path observables having different eigenstates $\left|\psi_{3}\right\rangle$ and $\left|\psi_{4}\right\rangle$ with eigenvalues $\pm1$.

\subsection{The spin measurement and the argument inferring quantum contextuality} 
In our earlier paper\cite{pan}, we considered the measurement of an arbitrary spin variable, say, $\widehat\sigma_{\theta}$ by SG1 and SG2 (Fig.1)placed along the spatially separated channels $|\psi_{3}\rangle$ and $|\psi_{4}\rangle$ respectively, 
where 
\begin{equation}
\widehat{\sigma}_{\theta}= \left(\begin{array}{cl} 
    cos 2\theta& \ sin 2\theta\\  sin 2\theta& \  -cos 2\theta \end{array}\right) 
\end{equation}
Then, corresponding to the prepared path-spin entangled state $\left|\Psi\right\rangle_{PH}$ given by Eq.(1), the expectation value of the spin variable $\widehat\sigma_{\theta}$ as measured by considering the \emph{whole ensemble} of particles emerging from the beam-splitter BS2(part of the measuring setup here) is of the form

\begin{equation}
\langle\widehat{\sigma}_{\theta}\rangle_{\Psi}=(\bar{\sigma}_{\theta})_{SG1}+ (\bar{\sigma}_{\theta})_{SG2}=0
\end{equation}

Now, the measurement of $\widehat\sigma_{\theta}$ involves contributions from both the output subensembles  corresponding to the counts \emph{separately} registered in the measuring devices SG1 and SG2. The respective \emph{subensemble spin mean values} denoted by $(\bar{\sigma}_{\theta})_{SG1}$ and $(\bar{\sigma}_{\theta})_{SG2}$ are given by

\begin{subequations}
\begin{eqnarray}
&&(\bar{\sigma}_{\theta})_{SG1}=\frac{1}{2}(\delta^{2}-\gamma^{2}) cos 2\theta + \gamma\delta sin2\theta\\
\nonumber
\\
&&(\bar{\sigma}_{\theta})_{SG2}=\frac{1}{2}(\gamma^{2}-\delta^{2})cos2\theta - \gamma\delta sin2\theta
\end{eqnarray}
\end{subequations}
Now, in order to inferring quantum contextuality, we noted that the spin expectation value $\langle\widehat{\sigma}_{\theta}\rangle_\Psi$ pertaining to the whole ensemble, is \emph{independent} of $\gamma(\delta)$ and thus insensitive to the choices of the `path' observable that is measured along with it. But, each of the subensemble spin mean values $(\bar{\sigma}_{\theta})_{SG1} $ and $(\bar{\sigma}_{\theta})_{SG2} $ given by Eqs.(6a) and (6b) are \emph{contingent upon} the parameter $\gamma(\delta)$, and thus \emph{depend} on the context in which they have been measured. This is the precise form of path-spin `contextuality' we had demonstrated in our earlier paper that holds good \emph{whatever} for an arbitrary spin variable $\widehat{\sigma}_{\theta}$. In our paper\cite{pan}, we had also mentioned that this quantum mechanical effect of path-spin interdependence cannot be reproduced by a realist model for four dimensional system using pre-existing noncontextual values of dynamical variables. 

Now, we shall briefly summarize the essence of the comment on our paper by De. Zela\cite{dezela}.
\section{The setup of De Zela}

\begin{figure}[h]
{\rotatebox{0}{\resizebox{9.0cm}{7.0cm}{\includegraphics{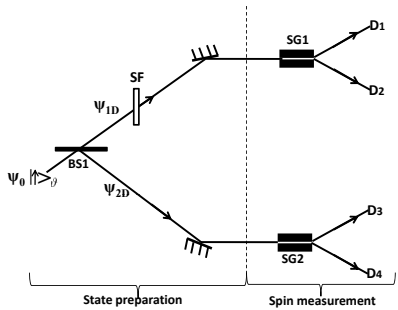}}}}
\vskip -1.2cm
\caption{\footnotesize An alternative setup used by De Zela\cite{dezela} comprising a 50:50 beam-splitter(BS), spin-flipper(SF) and  two spatially separated Stern-Gerlach devices SG1 and SG2. This setup has two  stages in contrast to our three distinct stages are the following: (i) The state preparation -  the input state $\left|\uparrow\right\rangle_{\vartheta}$ that incidents on BS passes through channels corresponding to  $|\psi_{1}\rangle_{D}$ and $|\psi_{2}\rangle_{D}$ and the SF is placed along one of the channels $|\psi_{1}\rangle_{D}$ that flips $\left|\uparrow\right\rangle_{\vartheta}$ to $\left|\downarrow\right\rangle_{\vartheta}$ thereby preparing an path-spin entangled state  (ii) The spin measurement - same as given in Fig.(1). Note that, no path measurement has been performed.}
\end{figure}
In his argument, De Zela has claimed that our proposed experimental setup(Fig.1) is a variant of SG setup(Fig.2). Let us closely examine his setup (Fig.2) which is redrawn by specifically marking the stages involving the state preparation and the spin measurement in that setup. In fact, in the setup considered by De Zela's (his Fig.2), no path measurement occurs along with spin measurement. What he has considered is effectively an extended state preparation (as compared to our setup) that is done by incorporating our path measurement within the sector of his state preparation. As a consequence, in the experimental arrangement discussed by De Zela, \emph{no} different context(concomitant path measurement) can be set while considering measurement of the spin observable $\widehat{\sigma}_{\theta}$. In particular, it needs to be stressed that if the state preparation is made in this way, the path degree of freedom has \emph{no} role thereby making the measurement simply restricted to two dimensional spin system, and then as is well known, this result can be reproduced by the KS model for a single qubit. This is really the essence of De Zela comment on our paper. But, its irrelevance to our paper lies in the way his setup differs from what we have considered in our demonstration. 
\subsection{De Zela's extended state preparation}
In Ref.\cite{dezela}, De Zela considered an ensemble of neutrons having an input state $\left|\uparrow\right\rangle_{\vartheta}= sin\vartheta\left|\uparrow\right\rangle_{z}+cos\vartheta\left|\downarrow\right\rangle_{z}$(relacing our $\left|\uparrow\right\rangle_{z}$) incident on a 50:50 beam splitter(Fig.2) that has two exit channels $|\psi\rangle_{1D}$ and $|\psi\rangle_{2D}$. In the upper channel, a spin-flipper is placed that flips $\left|\uparrow\right\rangle_{\vartheta}$ to $\left|\downarrow\right\rangle_{\vartheta}$. Then, the state(denoted  by $\left|\Phi\right\rangle_{D}$ instead of $\left|\Phi\right\rangle$ given by Eq.(9) in Ref.\cite{dezela}) of the neutrons after emerging from beam-splitter and spin-flipper is given by

\begin{eqnarray}
\left|\Phi\right\rangle_{D}= \frac{1}{\sqrt{2}}\left[\left|\psi_{1}\right\rangle_{D}\left|\downarrow\right\rangle_{\vartheta} +  i \left|\psi_{2}\right\rangle_{D}\left|\uparrow\right\rangle_{\vartheta}\right]
\end{eqnarray}
This is the prepared state on which De Zela has considered the spin measurement. In contrast, our prepared state $|\Psi\rangle_{PH}$ is given by Eq.(1). De Zela has remarked that the only difference between $|\Psi\rangle_{PH}$ and $|\Phi\rangle_{D}$ is a change in spin orientation and any statement addressing the kind of `contextuality' introduced by us by referring to $|\Psi\rangle_{PH}$ can also be made by referring $|\Phi\rangle_{D}$. 
With reference to his specific setup(Fig.2)this statement is completely misleading in confusing the state preparation and the relevant measurements. 
We stress here that if the path measurement is included as part of state preparation procedure, the argument for showing path-spin contextuality does not get off the ground. That is what exactly happened in De Zela's scheme. In contrast, the goal of our paper was to demonstrate path-spin interdependence. Most importantly, in order to show path-spin contextuality, one has to first carefully specify the context of measuring the path observable. This is, in fact, a general requirement in any test of noncontextual realist model involving two different degrees of freedom of a single particle; see, for example, Ref.[3-5]. 

\subsection{De Zela's spin measurement results and argument of its reproducibility by KS model for single qubit}

Now, on the prepared state $|\Phi\rangle_{D}$ given by Eq(7), De Zela considered the measurement of given  spin variable $\widehat{\sigma}_{\theta}$ along two different channels $|\psi_{1D}\rangle$ and $|\psi_{2D}\rangle$. The subensemble spin mean values are then the same as our results given by Eqs.(6a) and (6b), if one puts $\gamma=sin \vartheta$ and  $\delta=cos \vartheta$. In this way the results of our demonstration can be reproduced by using a variant of SG setup where the path measurement is included as  part of the state preparation procedure.

But what does this demonstration mean? De Zela has claimed that changing of the parameter $\vartheta$ is similar to changing $\gamma(\delta)$ in our scheme. This is simply a fallacious contention because the parameter $\vartheta$ pertains to the spin degree of freedom, while the parameter $\gamma(\delta)$ in our scheme determines the context of measuring the path observable. Further, it is not surprising that for a single qubit system, the  different state preparations may produce different subensemble mean values by preserving the expectation value. But, to stress once again, this is \emph{wholly irrelevant} to the entire issue of contextuality.  

\section{Conclusions}
To summarise, we have countered  De Zela's claim\cite{dezela} that our setup for demonstrating path-spin contextuality is simply a variant of the SG setup by showing that his setup(Fig.2) is \emph{not} at all equivalent to our setup(Fig.1), and hence his subsequent discussion of the reproducibility of our results by the KS realist model is irrelevant to not only our work but also the basic issue of contextuality. It is one's \emph{choice} what feature one would like to demonstrate in a given experiment. If one includes the path measurement within  an extended state preparation, then for such an experimental procedure, it would be pointless to discuss \emph{any form} of path-spin contextuality because there would be no scope for specifying/varying the choice of the path observable(that fixes the `context')for a given spin measurement. 

Further, we may add that there is an extended literature, both theoretical and experimental, related to the issue of testing noncontextual realist models in the four dimensional state space using the path and the spin degrees of freedom of a single particle; for example, Hasewawa \emph{et al.}\cite{hasegawa1,hasegawa2}, Bartosik \emph{et al.}\cite{hasegawa3}, Kirchmair \emph{et al.}\cite{nature} and Amselem \emph{et al.}\cite{ams}. If one wishes to apply the argument of  De Zela\cite{dezela} to those experiments too by including the path measurement as part of an extended state preparation, then again no context in terms of the path measurement can be specified/varied for a given spin measurement; and, consequently,  \emph{no} meaningful discussion of the issue of quantum violation of path-spin noncontextuality is possible.

Finally, we may stress that if one aims to analyse our specific results in terms of realist models, first one should consider a four dimensional state space, and then an appropriate noncontextual realist model has to be formulated. Without doing this, merely by incorporating the path measurement as part of an extended state preparation thereby reducing the problem to the two dimensional state space, it is grossly misleading to make any claim regarding path-spin contextuality.       

\section*{Acknowledgments}
AKP acknowledges the Research Associateship of Bose Institute, Kolkata.  DH thanks the DST, Govt. of India and Centre for Science and Consciousness, Kolkata for support.

\end{document}